\documentclass[useAMS,usenatbib]{mn2e}
\usepackage{graphicx}
\usepackage{amsmath}
\usepackage{amsfonts}
\usepackage{amssymb}

\title[ExoFit: orbital parameters of extra-solar planets ]{\emph{ExoFit}: orbital parameters
of extra-solar planets from radial velocities}

\author [S.T. Balan and O. Lahav]{Sreekumar T. Balan$^{1,2}$\thanks{E-mail: st452@mrao.cam.ac.uk} and Ofer Lahav$^{1}$\thanks{E-mail: lahav@star.ucl.ac.uk}\\
$^{1}$Department of Physics and Astronomy,University College London, Gower Street, London, WC1E 6BT, UK\\
$^{2}$Astrophysics Group, Cavendish Laboratory, JJ Thomson Avenue, Cambridge, CB3 0HE, UK}

\begin{document}

\date{Accepted 2008 December 10. Received 2008 December 9; in original form 2008 May 21}
\pagerange{\pageref{firstpage}--\pageref{lastpage}} \pubyear{2008}
 \maketitle
\label{firstpage}

\begin{abstract}
 Retrieval of orbital parameters of extrasolar planets poses
 considerable statistical challenges.  Due to sparse sampling,
 measurement errors, parameters degeneracy and modelling limitations
 there are no unique values of basic parameters such as period and
 eccentricity.  Here we estimate the orbital parameters from radial
 velocity data in a Bayesian framework by utilising Markov Chain Monte
 Carlo (MCMC) simulations with the Metropolis-Hastings algorithm. We
 follow a methodology recently proposed by Gregory and Ford.
Our implementation of MCMC is based on 
 the object oriented approach outlined by Graves.
 We make our resulting code, \emph{ExoFit}, publicly
 available with this paper. It can search for either one or two planets as illustrated on mock data.
 As an example we re-analysed the orbital
 solution of companions to HD 187085 and HD 159868 from the published radial velocity
 data. 
 We confirm the degeneracy reported for orbital parameters of
 the companion to
 HD 187085 and show that a low eccentricity orbit is more probable for
 this planet. For HD 159868 we obtained slightly different orbital
 solution and a relatively high `noise' factor indicating the presence
 of an unaccounted signal in the radial velocity data. \emph{ExoFit}
 is designed in such a way that it can be extended for a variety of
 probability models, including different Bayesian priors. 

\end{abstract}

\begin{keywords}
stars:planetary systems - stars: individual: HD 187085, HD 159868 - techniques: radial velocities - methods: data analysis - methods: statistical - methods: numerical.
\end{keywords}

\section{Introduction}
The study of extra-solar planets is of great importance because it
helps us to understand the origin and evolution of the Solar System. The
information gained from analysing new stellar systems serves as a sign
post towards finding possible life forms outside our Solar System. 
The first 200 or so detected extra-solar planets has been discovered
using the radial velocity method. Other detection methods
include astrometric method, transit photometry, gravitational
micro-lensing and direct imaging.  Astrometric method look for the
periodic shifts in the position of star and transit method measures
the attenuation of star light caused by the passage of the planet
across the star. Gravitational micro-lensing utilises the
amplification of light rays from the background object by a an
intervening massive object.  In a number of cases,
existence of the planet has been confirmed by more than one method. The number of
multi-planet systems discovered has also increased during the past few
years due the improved precision of radial velocity measurements.

In this article, we are
concerned with the extraction of orbital parameters from observed
radial velocity data.
A survey of the literature in this field indicates that orbital parameters
and their uncertainties were traditionally obtained by a two step
method. Searching for periodicity in the radial velocity data using a
Lomb-Scargle~ \citep{lomb1973,scargle1982} periodogram to fix the
orbital period and then estimating other parameters by
Levenberg-Marquardt~\citep{levenberg1944,marquardt1964}
minimisation. Recently, Bayesian methods have been
applied~\citep{gregory2005a,ford2005,gregory2007} to find out the best fit orbital
parameters and these studies suggest that Bayesian techniques have
considerable advantages over the traditional methods; for e.g when the
data does not cover a single orbital phase of the planet. In this work we analyse the radial velocity data from a Bayesian
point of view and extract the orbital parameters and corresponding
uncertainties using Markov Chain Monte Carlo (MCMC) simulations.  We
make our code, \emph{ExoFit}, and a detailed documentation available on-line\footnote{http://www.star.ucl.ac.uk/$\sim$lahav/exofit.html}.

The outline of the paper is as follows.
In Section \ref{sec:modelling_rv} we summarise the modelling of 
radial velocities. In Section \ref{sec:bayesian_retrieval} 
we discuss the Bayesian approach to the problem.
In Section \ref{sec:mcmc_impl} we present the MCMC implementation
and the \emph{ExoFit} software package.
\emph{ExoFit} is then applied to mock and observed data 
in Sections \ref{sec:mock_data} and \ref{sec:observed_data} respectively. The results are discussed 
in Section \ref{sec:discussion}.
\section{Modelling of Radial Velocity}
\label{sec:modelling_rv}
\subsection{Doppler Spectrography}
Planets are many times fainter than their host stars because they shine only by reflecting the star light. This makes their direct imaging extremely difficult. However, the gravitational pull of the planet makes the star wobble and this produces measurable periodic shifts in the apparent speed of the parent star. The motion of the star around the centre of mass causes the observed spectrum of the star to be Doppler shifted according its radial velocity, i.e. the velocity along the line of sight of the observer. This is measured over a course of time to obtain the radial velocity data along with the measurement uncertainties.
\subsection{Radial Velocity of Star}
\label{sec:rv}
A single planet model is assumed here to analyse the radial velocity data. 
The radial velocity of a star can be written as~\citep{murray,suto}
\begin{eqnarray}
v_i &=&V-K\big(\sin(f_i+\varpi)+e\,\sin \varpi \big)\\
K &=& \frac{m_p}{m_s+m_p}\frac{na\,\sin i}{\sqrt{1-e^2}}\nonumber
\label{full_expression}
\end{eqnarray}
where $v_i$ is the $i$th radial velocity entry corresponding to time coordinate $t_i$ and,\\
$V=$ the systematic velocity of the system,\\
$m_p=$ the mass of the planet,\\
$m_s=$ the mass of the star,\\
$n=\frac{2\pi}{T}$ the mean motion and $T$ is orbital period of planet,\\
$a=$ the length of the semi-major axis of the planet,\\
$i=$ the inclination of the orbital plane with the ecliptic,\\
$e=$ the eccentricity of the planet,\\
$f_i=$ the true anomaly at time $t_i$ and\\
$\varpi=$ the longitude of periastron.\\
The radial velocity depends on time via the true anomaly $f_i$. 
The full formalism and comparison with a common approximation is discussed in the User's Guide to \emph{ExoFit}.
\subsection{Radial velocity data}
According to (http://exoplanet.eu) eighteen different radial velocity search programmes are looking for extrasolar planets. Majority of the contributions come from Keck, Lick and Anglo-Australian observatories (the California \& Carnegie and Anglo-Australian planet searches) and searches based at l'Observatoire de Haute Provence 	and La  Silla Observatory (the Geneva extrasolar planet search). Radial velocity data for a star consists of time of observation $t_i$, measured radial velocity $v_i$ and uncertainty associated with each measurement $e_i$. These uncertainties are a characteristic of the instruments used for measurements. The precision of these instruments have improved from the order of $10ms^{-1}$in 1994 to order of $1ms^{-1}$~\citep{butler2006a,mayor2004a} at present.\footnote{The measurement method and its uncertainties are discussed in the corresponding planet discovery papers.} This is extremely significant for finding low mass companions as well as planets with large $a$ s.
\section{Bayesian Retrieval of orbital parameters}
\label{sec:bayesian_retrieval}
\subsection{Introduction}
The extraction of orbital parameters from the radial velocity data poses considerable statistical challenges. 
Earlier in this article we mentioned the traditional two step method that is generally used to retrieve orbital parameters.
Studies by~\cite{cumming2004} and~\cite{cumming1999a} have identified two cases where these methods become inefficient in accurately characterising the orbital elements:
\begin{enumerate}
 \item When the orbital period is extremely short and the eccentricity is high.
\item When the duration of observation does not span at least a single orbital phase.
\end{enumerate}

Incomplete radial
velocity data gives rise to a multitude of orbital solutions which is
referred to as parameter degeneracy. Since the transit probability of the planet increases for short
periods, the orbital parameters predicted by the periodogram method can
be verified, in some cases, with the help of transit photometry. 
Higher eccentricities make the radial velocity curve less sinusoidal. \cite{OToole2007} makes use of 
a 2DKLS periodogram to incorporate the effect of eccentricity of the orbits while searching for orbital periods.
Recently Bayesian techniques have
been employed by~\cite{gregory2005a}, ~\cite{ford2005} and ~\cite{gregory2007}
 to retrieve
the orbital parameters of extra-solar planets. The results show that
Bayesian methods tackle the difficulties associated with the
traditional methods efficiently and transparently.
We emphasise that the relative merit
of different methods depends on the quality of the data.
Broadly speaking the choice of prior distribution may change the inference (posterior distribution) significantly when the quality of data is poor.
\subsection{The Bayesian method}
The starting point of any Bayesian analysis is Bayes'
theorem~\citep{bayes1763}. Let
$\mathbf{y}=(y_1,\ldots,y_i,\ldots,y_n)$ be a vector of $n$
observations whose probability distribution
$p(\mathbf{y}|\boldsymbol{\theta},H)$ is conditional on $k$ parameters
$\boldsymbol{\theta}=(\theta_1,\ldots,\theta_i,\ldots,\theta_k)$,
where $H$ represents the background information or the hypothesis by
which the probability statements are made. Suppose that the parameter
$\boldsymbol{\theta}$ has the probability distribution
$p(\mathbf{\theta}|H)$. Then, Bayes' theorem says
\begin{equation}
p(\boldsymbol{\theta}|\mathbf{y},H)=\frac{p(\mathbf{y}|\boldsymbol{\theta},H)\,p(\mathbf{\theta}|H)}{p(\mathbf{y}|H)}\,.
\label{eqn:bayes1}
\end{equation}
For a continuous $\boldsymbol{\theta}$ we can write
\begin{equation}
p(\mathbf{y}|H)=\int\,p(\mathbf{y}|\boldsymbol{\theta},H)\,p(\boldsymbol{\theta}|H)\,d\boldsymbol{\theta}\, ,
\end{equation}
which is constant for given $\mathbf{y}$ and a probability distribution $p(\boldsymbol{\theta}|H)$. Then Equation~[\ref{eqn:bayes1}] can be rewritten as
\begin{equation}
p(\boldsymbol{\theta}|\mathbf{y},H)=C\,p(\mathbf{y}|\boldsymbol{\theta},H)\,p(\boldsymbol{\theta}|H)\,.
\label{eqn:bayes_C}
\end{equation}
In the above equation $p(\boldsymbol{\theta}|H)$ is called \emph{prior distribution} of $\boldsymbol\theta$ since it conveys our knowledge about $\boldsymbol\theta$ before the data has been observed. Correspondingly, $p(\boldsymbol{\theta}|\mathbf{y},H)$ is known as the \emph{posterior distribution} of $\boldsymbol{\theta}$ given $\mathbf{y}$. The factor  $C$ is a normalising constant which ensures that the posterior distribution integrates to one. We call $p(\mathbf{y}|\boldsymbol{\theta},H)$ the likelihood function of $\boldsymbol\theta$ since $p(\mathbf{y}|\boldsymbol{\theta},H)$ can be considered as a function of $\boldsymbol\theta$ instead of $\mathbf{y}$. Then,
\begin{equation}
p(\boldsymbol{\theta}|\mathbf{y},H)\,\propto \,p(\mathbf{y}|\boldsymbol{\theta},H)\,p(\boldsymbol{\theta}|H)\,.
\end{equation}
Statistical inferences regarding $\boldsymbol{\theta}$ are derived from the posterior distribution of $\boldsymbol{\theta}$. The posterior distribution encapsulates all information about unknown quantities $\boldsymbol{\theta}$ following the observation of the data $\mathbf{y}$.
 The principal steps in the Bayesian method can be 
involve the Likelihood
$p(\textbf{y}|\boldsymbol{\theta},H)$, 
the prior 
$p(\boldsymbol{\theta}|H)$,
the posterior  $p(\boldsymbol{\theta}|\mathbf{y},H)$, 
and the resulting inference, e.g. ~\cite{ohagan2004}.

\subsection{Likelihood function}
\label{sec:lik_fun}
Let $d_i$ represent the measured radial velocity data for the $i$th instant of time $t_i$. Observed radial velocity data can be modelled by the equation~\citep{gregory2005a}
\begin{equation}
d_i=\nu_i+\epsilon_i+\delta\,,
\label{eqn:rv_model}
\end{equation}
where $\nu_i$ is the true radial velocity of the star and $\epsilon_i$ is the uncertainty component arising from accountable but
unequal measurement errors which are assumed to be normally
distributed. The term $\delta$ explains any unknown measurement
errors. There can be multiple reasons for the presence of this
uncertainty component~\citep{butler2006a}. For example this could be
the result of another planet in the system or caused by the intrinsic
anomalies in the star spectrum due to the irregularities on the
surface of the star~\citep{mayor2004a,mayor2003a,bouchy2005a}. Thus
any noise component that cannot be modelled is described by the term
$\delta$. The probability distribution of $\delta$ 
is chosen to be a Gaussian distribution with finite variance
$s^2$. Therefore the combination of uncertainties $\epsilon_i+\delta$ has a
Gaussian distribution with a variance equal to $\sigma_i^2+s^2$.

The true radial velocity $\nu_i$ is modelled by the  equation~(\ref{full_expression}). Thus the radial velocity $v_i$ predicted by the mathematical model at an instant $t_i$ is
\begin{displaymath}
v_i=V-K\,\big(\sin(f_i+\varpi)+e\,\sin \varpi \big)\,.
\end{displaymath}
Six model parameters namely $T,K,V,e,w,$and $\chi$, as defined in the section~(\ref{sec:rv}) are used to fit the above equation onto a given radial velocity data.

Each error term $\epsilon_i$ in equation (\ref{eqn:rv_model}) is independent. Since they are assumed to follow a Gaussian distribution, the likelihood function is product of $N$ Gaussians~\citep{gregory2005,gregory2005a} where $N$ is the number of observations. Thus
\begin{equation}
p(\mathbf{y}|\boldsymbol{\theta})=A\, \exp \Biggl[-\sum_{i=1}^{N}\frac{(d_i-v_i)^2}{2(\sigma_i^2+s^2)}\Biggr]\,,
\label{eqn:lik}
\end{equation}
where
\begin{equation}
A=(2\pi)^{-N/2}\Biggl[\prod_{i=1}^{N}\big(\sigma_i^2+s^2\big)^{-1/2}\Biggr]\,.
\label{eqn:likA}
\end{equation}
and $s$ becomes the seventh parameter in our probability model.

\subsection{Choice of Priors}
\label{sec:prior}
The choice of priors is extremely important in the Bayesian analysis
as senseless choice of priors can produce to misleading results. 
Physical and geometric conditions govern the selection of prior distributions for most of the parameters. Since $\boldsymbol{\theta}=(T,K,V,e,\varpi,\chi,s)$ the prior distribution in our problem can be written as
\begin{eqnarray}
p(\boldsymbol{\theta}|H)& =& p(T|H)\,p(K|H)\,p(V|H)\nonumber\\
&&\quad p(e|H)\,p(\varpi |H)\,p(\chi |H)\,p(s |H)\,,
\label{eqn:priors}
\end{eqnarray}
on the assumption that they are independent. We will discuss how the above conditions are met for our choice of prior for each parameter in the next few sections.

We follow the choice of priors as given by \cite{gregory2007},
as summarized in Table \ref{tab:prior}. Obviously, part of the Bayesian framework is the ability to change priors and to  check the sensitivity of the results to them. Our \emph{ExoFit} package allows this freedom. 
\begin{table}
\caption[Choice of Priors]
{The assumed  prior distribution of various parameters and their boundaries.
It is similar to choice of priors given by \cite{gregory2007}, except for
the prior distribution of $K$.}
\scriptsize
\begin{tabular}{lllllll}
\hline
Para. & Prior  & Mathematical Form & Min & Max \\ 
\hline
$T(days)$& Jeffreys & $\frac{1}{T\,\ln\Big(\frac{T_{max}}{T_{min}}\Big)}$ & 0.2 & 15000 \\ 
$K(ms^{-1})$ & Mod. Jeffreys & $\frac{(K+K_0)^{-1}}{\ln\big(\frac{K_0+K_{max}}{K_0})}$ & 0.0 &2000 \\ 
$V(ms^{-1})$ & Uniform  & $\frac{1}{V_{max}-V_{min}}$ & -2000 & 2000 \\ 
$e$ & Uniform & 1 & 0 & 1 \\ 
$\varpi$ & Uniform & $\frac{1}{2\pi}$ & 0 & $2\pi$ \\ 
$\chi$ & Uniform & 1 & 0 & 1 \\ 
$s(ms^{-1})$ & Mod. Jeffreys & $\frac{(s+s_0)^{-1}}{\ln\big(\frac{s_0+s_{max}}{s_0}\big)}$ & 0 & $2000$\\
\hline
\end{tabular}
\label{tab:prior}
\end{table}
\normalsize
\subsection{Posterior Distribution}
Posterior distribution is obtained by applying the Bayes' theorem
given by the equation~(\ref{eqn:bayes_C}). 
Useful and interesting features of the posterior
distribution should be identified before making summary
statements. For example, the posterior distribution may be unimodal
but asymmetric or it can be multi-modal with many probability peaks.
Any summary statistic(e.g. mean, median and mode) can be expressed in terms of
posterior expectations of
$\boldsymbol{\theta}$~\citep{gilks1996,berger1980}.
\section{MCMC Implementation}
\label{sec:mcmc_impl}
\subsection{The MCMC approach} 
\label{sec:MC}
Difficulty in evaluating the multi-dimensional integrals is an
inherent inability of any Bayesian formulation. Many techniques have
been developed in the last 25 years to deal with this problem.
Simulation methods dominate this area and
several computational algorithms were developed to numerically
integrate the posterior distribution in order to find out the marginal
distributions of each parameter. According to ~\cite{berg2004} the
abundance of computational power has produced a paradigm shift with
respect to statistics: Computationally intensive but conceptually
simple methods are preferred. Markov Chain Monte Carlo (MCMC) method
is one of the most commonly used methods for simulating complex
probability distributions\footnote{A general form this method given by
\emph{Metropolis-Hastings}~\citep{metropolis1953,hastings1970} algorithm
is explained in \emph{ExoFit User Guide}.}. Our code is based on the concepts outlined by
\cite{graves2007}. The emphasis is on the extensibility of the code to
accommodate different probability models to a certain extent.

Bayesian MCMC methods have gained popularity in various areas of
astrophysics, for example in multi-parameter estimation from
cosmological data sets (e.g. CosmoMC; ~\cite{lb2002}).  We emphasise
that the two ingredients, the Bayesian approach and the MCMC tool, are
distinct and not necessarily related.  One may apply MCMC to a
multi-parameter likelihood analysis (i.e. without priors-- equivalent
to a Bayesian method for uniform priors).  On the other hand one may
work out a full Bayesian method (with complicated priors) using a
grid-based maximisation procedure, without the need for MCMC. Having
said that, the combination of the Bayesian and MCMC methods is a very
powerful way to tackle our problem.

From a Bayesian point of view analysis of statistical problems
requires an efficient tool for simulating posterior densities and MCMC
methods are ideally suited for this purpose. In general, one must consider planet-planet interaction while modelling radial velocity of a star. In typical systems the radial
velocity of an $n$-planet model can be approximated as a linear combination
of $n$ single planet radial velocities.
With the present version of \emph{ExoFit}
it is possible to search for either one or two planets.

\subsection{The \emph{ExoFit} software package}
\emph{ExoFit} is a step towards achieving the goals mentioned
above. It should be considered as a platform to develop MCMC based
methods for estimating orbital parameters of a generalised
multi-planet model. Object oriented design of \emph{ExoFit} makes it
extremely well suited for extending the analysis to multi-planet systems with
prior constraints on several orbital parameters such as eccentricity
and length of semi-major axis. Following \cite{graves2007}, our
implementation MCMC consists of Data, State, Bond and Update. They are referred to as objects in object oriented analysis. \emph{data} handles the input data into the MCMC analysis. A \emph{state} consists of a set of parameters whose posterior distribution is sought. The parameter values at a particular instant defines the \emph{state} of Markov Chain in the analysis. The parameters in a particular \emph{state} are connected to each other by a \emph{bond}. It consists of prior densities and likelihood. For each state there corresponds a bond strength which is equal to $prior\times likelihood$. In other words it is the posterior density without the normalisation constant in Bayes' theorem. An \emph{update} selects the parameters that should be updated at particular iteration. New values for the parameters are proposed according to the \emph{update} defined and the new \emph{bond} strength is then calculated for the proposed state. The new state is accepted or rejected according Metropolis-Hastings method.

The central concept of this approach is that, the MCMC engine remains the same and need not be re-implemented whenever the probability model gets changed. We also take advantage of the commonalities among the different components of MCMC.
Our implementation works for variety or prior distributions ($Bonds$) and \emph{Update} methods. The only component that requires to be changed is the likelihood function. 
 
Convergence is an important aspect of any MCMC method and the choice of 
proposal distribution is absolutely crucial for achieving convergence
of the chain. Choosing an effective proposal distribution is very
difficult in many cases. \emph{ExoFit} employs an adaptive
Metropolis algorithm described by \cite{haario99} to fine tune the
proposal distribution step sizes.
\section{Application to mock data}
\label{sec:mock_data}
We used simulated data sets to test the accuracy of the orbital
parameters extracted by \emph{ExoFit}. The output of \emph{ExoFit} is
analysed with the help of the R statistical environment\footnote{http://www.r-project.org/}
which is freely available. A number of
packages are available in R for checking the convergence of Markov
Chains and creating the histograms and density plots of posterior
distribution of parameters. 
\begin{table*}
 \centering
\caption{Column 2 shows the simulated radial velocities obtained from a single planet model (Table \ref{tab:simcomp}) using equation (\ref{full_expression}). A Gaussian noise with $\mu=0.0ms^{-1}$ and $\sigma=2.0ms^{-1}$ has been added to each radial velocity term. Columns 3 is obtained by adding the radial velocity from a second planet (see Table \ref{tab:simcomp}) to each term in column 2, thus creating a mock radial velocity data for a two planet system. The measurement uncertainties (Gaussian distribution with $\mu=2.0ms^{-1}$ and $\sigma=0.5ms^{-1}$) are shown in columns 4 and 5.}
\begin{tabular}{llllll}
\hline
$JD(-2451000)$ & $RV(ms^{-1})$ & $Error(ms^{-1})$&  & $RV(ms^{-1})$ & $Error(ms^{-1})$\\ 
\hline
 & 1-planet &  &  &2-planets  &  \\
\hline
1400.15200	&	-17.90	&	2.1	&	&	-25.30	&	2.1	\\
1413.87749	&	-18.50	&	2.8	&	&	-18.20	&	2.8	\\
1434.46573	&	-30.00	&	1.4	&	&	-22.50	&	1.4	\\
1455.05396	&	-33.00	&	2.0	&	&	-26.20	&	2.0	\\
1489.36769	&	-41.10	&	2.1	&	&	-52.70	&	2.1	\\
1496.23043	&	-41.30	&	2.4	&	&	-50.80	&	2.4	\\
1503.09318	&	-45.50	&	1.7	&	&	-51.20	&	1.7	\\
1564.85788	&	-47.60	&	1.7	&	&	-45.30	&	1.7	\\
1571.72063	&	-44.20	&	2.0	&	&	-46.80	&	2.0	\\
1585.44612	&	-39.40	&	1.6	&	&	-50.70	&	1.6	\\
1592.30886	&	-33.60	&	1.9	&	&	-44.70	&	1.9	\\
1612.89710	&	-14.50	&	1.9	&	&	-14.70	&	1.9	\\
1619.75984	&	-6.17	&	1.9	&	&	-3.07	&	1.9	\\
1654.07357	&	38.20	&	2.4	&	&	45.30	&	2.4	\\
1688.38729	&	61.90	&	1.6	&	&	50.30	&	1.6	\\
1695.25004	&	64.90	&	1.7	&	&	54.90	&	1.7	\\
1715.83827	&	74.00	&	2.9	&	&	75.30	&	2.9	\\
1722.70102	&	73.70	&	2.7	&	&	78.00	&	2.7	\\
1736.42651	&	70.20	&	1.8	&	&	78.00	&	1.8	\\
1743.28925	&	72.70	&	2.6	&	&	81.10	&	2.6	\\
1770.74024	&	67.30	&	2.3	&	&	65.50	&	2.3	\\
1777.60298	&	65.10	&	2.0	&	&	57.80	&	2.0	\\
1784.46573	&	66.20	&	2.7	&	&	55.20	&	2.7	\\
1798.19122	&	61.90	&	2.4	&	&	53.40	&	2.4	\\
1811.91671	&	56.70	&	1.9	&	&	56.00	&	1.9	\\
1818.77945	&	56.20	&	2.2	&	&	58.80	&	2.2	\\
1832.50494	&	51.00	&	1.7	&	&	58.20	&	1.7	\\
1901.13239	&	32.80	&	1.6	&	&	25.90	&	1.6	\\
1907.99514	&	32.40	&	2.0	&	&	29.60	&	2.0	\\
1921.72063	&	29.80	&	2.2	&	&	33.70	&	2.2	\\
1928.58337	&	33.90	&	3.7	&	&	40.10	&	3.7	\\
1935.44612	&	29.10	&	2.9	&	&	36.80	&	2.9	\\
1962.89710	&	19.00	&	2.2	&	&	22.40	&	2.2	\\
1969.75984	&	17.40	&	2.2	&	&	16.30	&	2.2	\\
\hline
 \end{tabular}
 \label{tab:simdata}
\end{table*}
\begin{table*}
 \centering
\scriptsize
 \caption{A comparison between input parameters to the mock data for 1 and 2 planets and
the parameters extracted with \emph{ExoFit} 1-planet model. Columns 3, 4 and 5 show the
posterior mean (and standard deviation), median (and 25\% and 75\%
quantiles) and the maximum a posteriori, i.e. the posterior mode (and 68.3\% highest density
regions) obtained by the application of $ExoFit$ 1-planet model to the single planet data from Table \ref{tab:simdata}. Column 6 shows the orbital parameters of the second planet used to simulate the data for the two planet system (columns 4 and 5 in Table \ref{tab:simdata}). Summary statistics of the posterior distribution generated with $ExoFit$ 1-planet model for this 2-planet data is shown in columns 7, 8 and 9. Note the difference between the noise factors extracted with $ExoFit$ from the single planet data and the 2-planet data. Thus the noise factor $s$ servers as an indicator for a possible embedded signal in the radial velocity data apart from the random Gaussian errors.}
\begin{tabular}{@{}lllllllll@{}}
\hline
Parameter & Planet 1 &  \emph{ExoFit} & \emph{ExoFit} & \emph{ExoFit}  & Planet 2 & \emph{ExoFit}  & \emph{ExoFit} & \emph{ExoFit} \\
 & & Mean & Median & Mode & &  Mean & Median & Mode \\
\hline
$V (ms^{-1})$ & 12.0 & $11.91\pm 0.44$ & $ 11.91^{+0.30}_{-0.30}$ & $11.91^{+0.44}_{-0.45}$& & $11.50\pm1.60$& $11.55^{+1.00}_{-1.05}$& $11.69^{+1.43}_{-1.64}$\\
\hline
$T(days)$ & 700.00 & $ 704.85\pm12.91$ & $704.24^{+9.03}_{-8.36}$ & $702.79^{+13.62}_{-11.67}$ & $100.0$ & $719.11\pm53.06$ &$711.23^{+33.62}_{27.52}$& $704.95^{+44.76}_{-43.50}$\\
\hline
$K (ms^{-1})$ & 60.00& $ 60.39\pm  0.56$ & $60.39^{+0.38}_{-0.38}$ & $60.37^{+0.58}_{-0.54}$& $10.0$& $62.52\pm2.25$ & $62.51^{+1.44}_{-1.46}$& $62.48^{+2.19}_{-2.20}$ \\
\hline
$e$ & 0.38 & $ 0.38\pm 0.03$ & $ 0.38^{+0.01}_{-0.01}$ & $0.38^{+0.01}_{-0.02}$& $0.18$& $0.40\pm0.04$& $0.40^{+0.03}_{-0.03}$& $0.40^{+0.04}_{0.04}$ \\
\hline
$\varpi$ & 3.10 & $3.09\pm0.03$ & $3.09^{+0.02}_{-0.02}$ & $3.04^{+0.02}_{-0.03}$ &$1.10$& $ 2.96\pm 0.11$ & $ 2.96^{+0.07}_{-0.07}$& $ 2.96^{+0.11}_{-0.11}$\\
\hline
$\chi$ & 0.67 & $ 0.67\pm0.00$ & $0.67^{+0.00}_{-0.01}$ & $0.67^{+0.00}_{-0.01}$&$0.17$& $0.69\pm0.01$ & $ 0.69^{+0.01}_{-0.01}$ & $0.68^{+0.01}_{-0.01}$\\
\hline
$s$ &  & $0.4276\pm0.35$ &  $ 0.34^{+0.27}_{-0.19}$ & $0.09^{+0.43}_{-0.09}$& & $7.49\pm1.14$ & $7.36^{+0.80}_{-0.66}$ & $7.12^{+0.93}_{-0.93}$ \\
\hline
 \end{tabular}
 \label{tab:simcomp}
\end{table*}
\subsection{1-planet data with 1-planet fit}
\begin{figure}
 \centering
\includegraphics[scale=0.6]{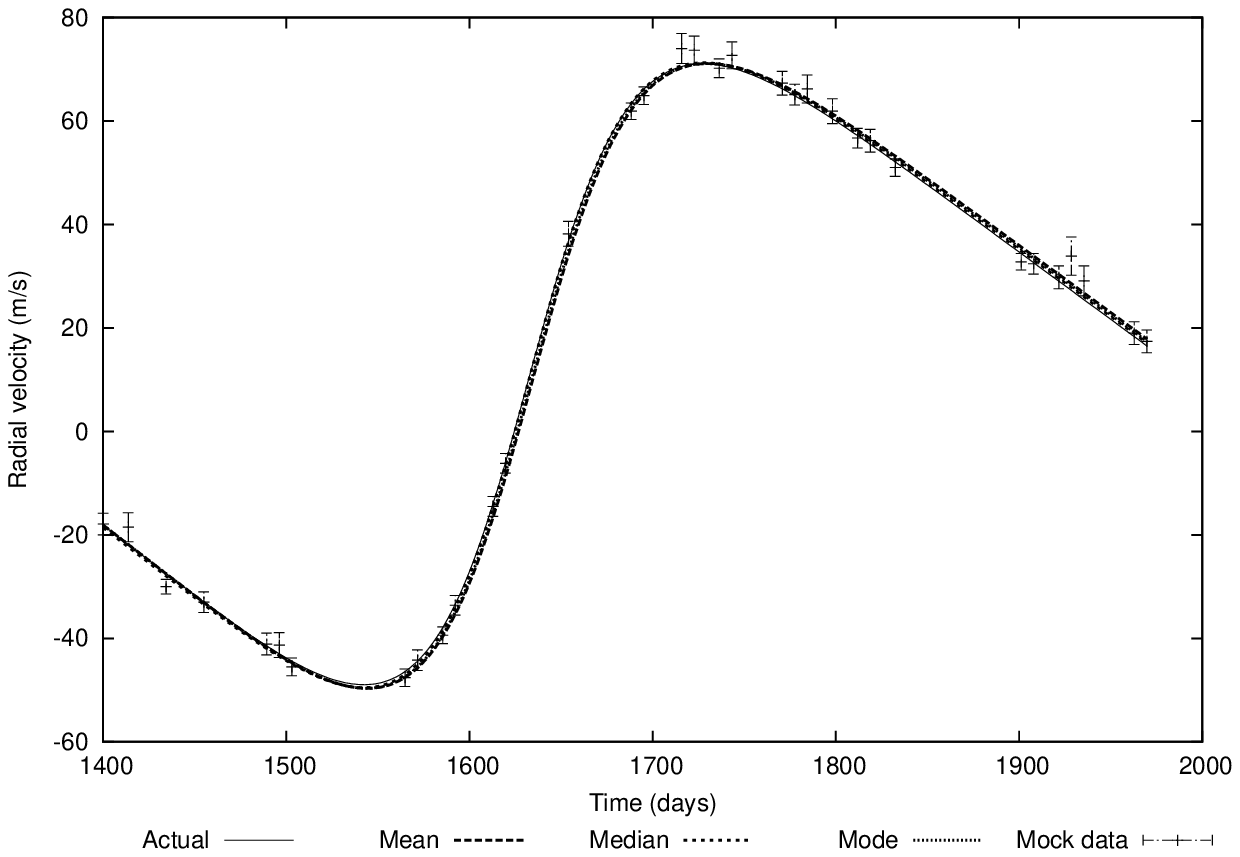}
 \caption{Radial velocity curves created with mean, median and mode of the posterior distribution of parameters. The radial velocity curve for actual input parameters with the simulated data is also shown.}
\label{fig:sim1}
 \includegraphics[scale=0.6]{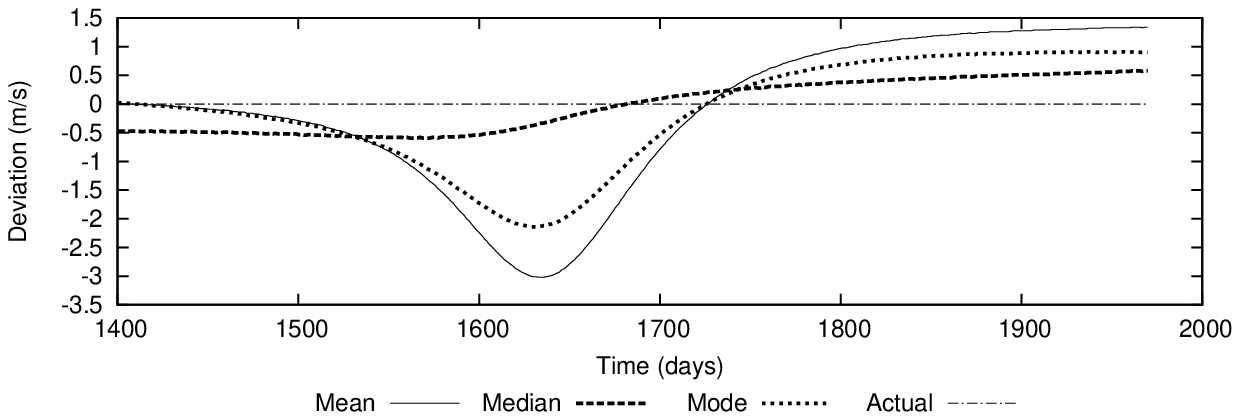}
 \caption{Deviation from the actual radial velocities of the simulated data for 3
 different summary statistics shown in Table~\ref{tab:simcomp}. Note
 the differences are no more than $3 ms^{-1}$.}
 \label{fig:simdvn}
\end{figure}

In this 
The first set of radial velocity entries as shown in Table~\ref{tab:simdata} (columns 1, 2 and 3) is created using a single planet model (equation (\ref{full_expression})) and adding a Gaussian noise. 
Table~\ref{tab:simcomp} shows the input
parameters used to create the data set and the parameters obtained
with \emph{ExoFit} for a single planet search.
Three posterior summary statistics are
compared in Table~\ref{tab:simcomp}; where  column 2 gives the assumed input
values, column 3 contains the mean values of MCMC samples for
each parameter along with the sample standard deviation, column 4
shows the median values with 75\% and 25\% quantiles. We used
\cite{hyndman1996}'s method to calculate \emph{Highest Density
Regions} (HDRs) of the posterior distribution. Column 5 contains
posterior modes (maximum a posteriori) 
and associated 68.3\% credible regions.
The difference between
actual radial velocity curve and the radial velocity curve for the
parameters obtained from \emph{ExoFit} is shown in the Fig.~\ref{fig:sim1}. We also plot the deviation of the radial velocities
from the actual one for each of these orbital solutions in
Fig.~\ref{fig:simdvn}.
We note the deviations between the estimators are smaller than $3 ms^{-1}$. The noise factor $s\approx 0.4$ is less than the measurement uncertainties, implying the assumed radial velocity model is ``good'' and there is no sign of an additional signal present in the data.
\subsection{2-planet data with 1-planet fit}
\begin{figure}
 \centering
\includegraphics[scale=0.6]{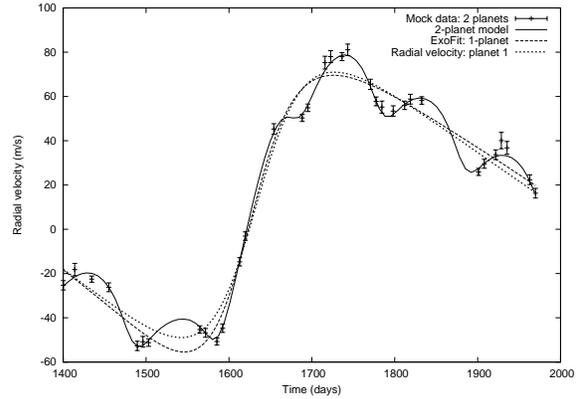}
\caption{The radial velocity curve for two planet model with orbital parameters of the planets from Table \ref{tab:simcomp} along with the radial velocity plot created with median values of parameters obtained by $ExoFit$ (median values) and the actual radial velocity.}
\label{fig:rv_curves_2p}
\end{figure}

A second set of radial velocity entries as shown in column 4  of Table \ref{tab:simdata} has been created by adding the radial velocity from a second planet with orbital parameters as shown in Table \ref{tab:simcomp} to the radial velocity data of the single planet model in column 2 of Table \ref{tab:simdata}.
The results from the application of $ExoFit$ to this simulated data is shown in columns 7, 8 and 9 of Table \ref{tab:simcomp}. Noise factor $s\approx 7ms^{-1}$ in this case is clearly higher than the measurement uncertainties, indicating the presence of a signal unaccounted for in the radial velocity model. Note that the noise factor obtained in the previous case was only $\approx 0.4ms^{-1}$. Fig. \ref{fig:rv_curves_2p} shows the radial velocity curve for the median values of the posterior distribution of orbital parameters from $ExoFit$ along with the actual radial velocity plot for the 2-planet model. 
\subsection{A 2-planet fit to 2-planet mock data}
\label{sec:2planets}
$ExoFit$ has an option to search for two planets in the radial velocity data. We choose the probability model to be similar to that of the single planet model explained in Section \ref{sec:lik_fun}. The observed radial velocity data is again modelled by the equation \ref{eqn:rv_model}. The radial velocity $v_i$ predicted by the mathematical model at an instant $t_i$ is given by 
\begin{align}
 v_i = V &-\Big(\,K_1\big(\sin(f_{i1}+\varpi_1)+e_1\,\sin \varpi_1\big)\nonumber \\
& + K_2\big(\sin(f_{i2}+\varpi_2)+e_2\,\sin \varpi_2\big) \,\Big)\, .
\end{align}
11 parameters $\{V,T_1,K_1,e_1,w_1,\chi_1,T_2,K_2,e_2,w_2,\chi_2\}$\footnote{Subscripts 1 and 2 indicate planets 1 and 2 respectively.}, as defined in the Section \ref{sec:rv} are used to fit the above equation onto the radial velocity data. The likelihood function is again given by equations \ref{eqn:lik} and \ref{eqn:likA} respectively, and $s$ becomes the $12^{th}$ parameter in our probability model. The choice of prior distributions for each of these parameters is given in Table \ref{tab:prior_12}.
\begin{table}
\centering
\scriptsize
\caption[Choice of Priors for the 2-planet model]
{The assumed  prior distribution of orbital parameters and their boundaries for a 2-planet model. The boundaries for $K2$ can be made smaller in order to speed up the convergence of the Markov Chain.}
\begin{tabular}{lllllll}
\hline
 Para. & Prior  & Mathematical Form & Min & Max \\ 
\hline
$V_(ms^{-1})$ & Uniform  & $\frac{1}{V_max-V_{min}}$ & -2000 & 2000 \\ 
 $T_1(days)$& Jeffreys & $\frac{1}{T_1\,\ln\Big(\frac{T_{1\,max}}{T_{1\,min}}\Big)}$ & 0.2 & 15000 \\ 
$K_{1}(ms^{-1})$ & Mod. Jeffreys & $\frac{(K_1+K_{1\,0})^{-1}}{\ln\big(\frac{K_{1\,0}+K_{1\,max}}{K_{1\,0}})}$ & 0.0 &2000 \\ 
 $e_1$ & Uniform & 1 & 0 & 1 \\ 
 $\varpi_1$ & Uniform & $\frac{1}{2\pi}$ & 0 & $2\pi$ \\ 
 $\chi_1$ & Uniform & 1 & 0 & 1 \\ 
 $T_2(days)$& Jeffreys & $\frac{1}{T_2\,\ln\Big(\frac{T_{2\,max}}{T_{2\,min}}\Big)}$ & 0.2 & 15000 \\ 
 $K_{2}(ms^{-1})$ & Mod. Jeffreys &  $\frac{(K_2+K_{2\,0})^{-1}}{\ln\big(\frac{K_{2\,0}+K_{2\,max}}{K_{2\,0}})}$ & 0.0 &2000 \\ 
 $e_2$ & Uniform & 1 & 0 & 1 \\ 
$\varpi_2$ & Uniform & $\frac{1}{2\pi}$ & 0 & $2\pi$ \\ 
$\chi_2$ & Uniform & 1 & 0 & 1 \\ 
$s(ms^{-1})$ & Mod. Jeffreys & $\frac{(s+s_{0})^{-1}}{\ln\big(\frac{s_{0}+s_{max}}{s_{0}}\big)}$ & 0 & $2000$\\
\hline
\end{tabular}
\label{tab:prior_12}
\end{table}
\normalsize

\begin{figure}
 \centering
 \includegraphics[scale=0.6]{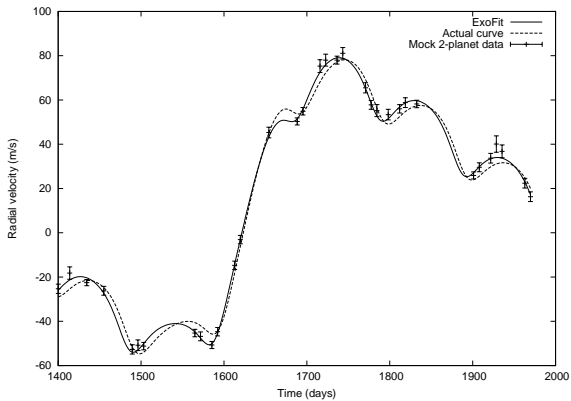}
 \caption{A comparison of the actual 2-planet radial velocity curve and the one obtained with \textit{ExoFit} (median values).}
 \label{fig:comp_res}
\end{figure}

\begin{figure}
 \centering
 \includegraphics[scale=0.4]{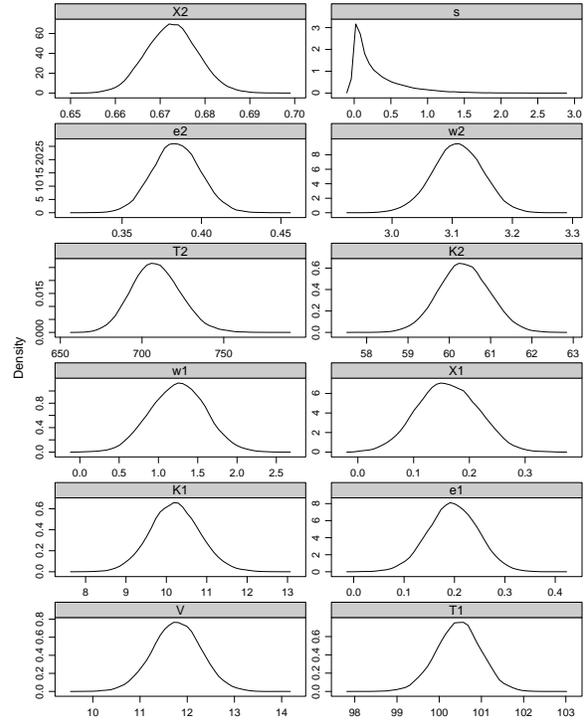}
 \caption{Posterior density of 12 orbital parameters of the 2-planet model obtained from $ExoFit$.}
 \label{fig:dens_12}
\end{figure}

We apply the $ExoFit$ to the 2-planet mock radial velocity data given in Table \ref{tab:simdata} (column 4). A comparison of actual orbital parameters with the ones extracted with $ExoFit$ is given in Table \ref{tab:results_exo_multi}. Fig \ref{fig:comp_res} shows the actual radial velocity curve and the one created with the median values of the posterior distribution of orbital parameters  obtained from $ExoFit$\footnote{We caution that, using median values for orbital parameters with skewed marginal posterior distributions can lead to models that very poor.}.Note that the noise factor $s$ is now $\approx0.4ms^{-1}$ compared to $s\approx7ms^{-1}$ in the single planet model. The density plots are shown in Fig \ref{fig:dens_12}.
\begin{table}
\label{tab:results_exo_multi}
\scriptsize
\caption{A comparison of actual values of orbital parameters used to create the mock radial velocity data for the 2-planet system (Columns 4 and 5 in Table \ref{tab:simdata}) and the summary of the posterior distribution of orbital parameters extracted with \textit{ExoFit} 2-planet model. Column 3, 4 and 5 shows
posterior mean (and standard deviation), median (and 25\% and 75\%
quantiles) and the maximum a posteriori, i.e. posterior mode (and 68.3\% highest density
regions) respectively.}
 \begin{tabular}{@{}lllll@{}}
\hline
Parameter & Actual & \emph{ExoFit} & \emph{ExoFit}& \emph{ExoFit} \\
 & & Mean & Median & Mode \\
\hline

$V (ms^{-1})$ & $12.0 $ & $11.80\pm 00.52$ & $ 11.80^{+0.33}_{-0.35}$ & $11.87^{+0.44}_{-0.58}$ \\
\hline
$T_1(days)$ & $700.0 $ & $  709.06\pm15.03$ & $ 708.25^{+10.32}_{-9.68}$ & $708.03^{+14.04}_{-15.36}$ \\
\hline
$K_1 (ms^{-1})$ & $60.0 $ & $ 60.34\pm 0.62$ & $60.33^{+0.41}_{-0.41}$ & $60.33^{+0.59}_{-0.64}$ \\
\hline
$e_1$ & $0.38 $ & $  0.38\pm 0.01$ & $ 0.38^{+0.01}_{-0.0}$ & $0.38^{+0.01}_{-0.01}$ \\
\hline
$\varpi_1$ & $3.10 $ & $3.10\pm 0.04$ & $3.10^{+0.02}_{-0.02}$ & $3.11^{+0.03}_{-0.04}$ \\
\hline
$\chi_1$ & $0.67 $ & $ 0.67\pm0.00$ & $0.67^{+0.00}_{-0.01}$ & $0.67^{+0.00}_{-0.00}$\\
\hline
$T_2(days)$ & $100.0 $ & $ 100.44\pm 0.53$ & $100.45^{+0.35}_{-0.36}$ & $100.43^{+0.54}_{-0.52}$ \\
\hline
$K_2 (ms^{-1})$ & $10.0 $& $ 10.19\pm0.62$ & $10.18^{+0.38}_{-0.41}$ & $10.14^{+0.65}_{-0.58}$ \\
\hline
$e_2$ & $0.18 $ & $  0.19\pm 0.05$ & $ 0.19^{+0.03}_{-0.03}$ & $0.20^{+0.05}_{-0.05}$ \\
\hline
$\varpi_2$ & $1.10 $ & $ 1.27\pm0.36$ & $1.28^{+0.23}_{-0.24}$ & $1.29^{+0.38}_{-0.38}$ \\
\hline
$\chi_2$ & $0,17 $ & $ 0.15\pm 0.05$ & $0.15^{+0.03}_{-0.03}$ & $0.15^{+0.05}_{-0.05}$\\
\hline
$s$ &  & $  0.50\pm0.41$ & $0.40^{+0.33}_{-0.22}$ & $0.15^{+0.05}_{-0.05}$\\
\hline
 \end{tabular}
\end{table}

Constraining the longitude of periastron $\varpi$ becomes increasingly difficult when the eccentricity of planetary system approaches zero. \cite{ford2006} suggests that, the efficiency of MCMC sampler in this situation can be increased by adopting a new parametrisation with $e\sin\,\varpi$ and $e\cos\,\varpi$. For discussions about further improvements in parametrisation, choice of priors and proposal distributions for MCMC, see \cite{ford2006}. These considerations will be implemented in the future version of \textit{ExoFit}. 
\section{Application to observed data}
\label{sec:observed_data}
In this section we apply \emph{ExoFit} to published radial velocity data. We choose data sets in which the measurement noise is relatively high and the entries are poorly sampled. The following examples shows that \emph{ExoFit} does a good job in estimating the posterior distribution orbital parameters.
\subsection{The companion to HD 187085}
\begin{figure}
 \centering
  \includegraphics[scale=0.6]{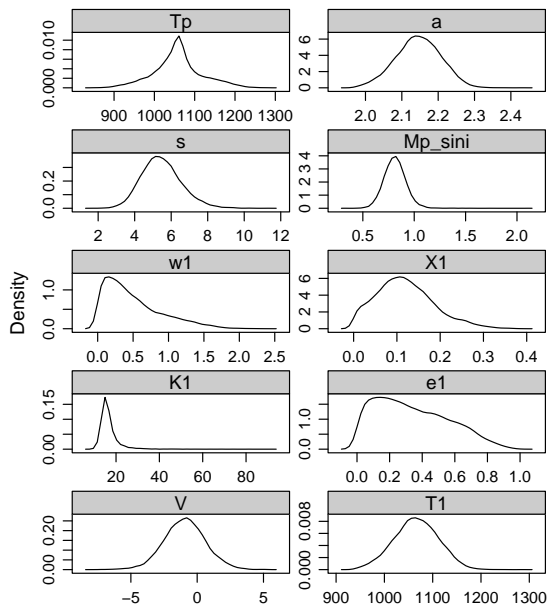}
 \caption{Posterior distribution of orbital parameters of the companion to HD 187085. These plots were created with R from the output of \emph{ExoFit}.}
 \label{fig:histHD187085}
\end{figure}

\begin{figure}
 \centering
 \includegraphics[scale=0.33]{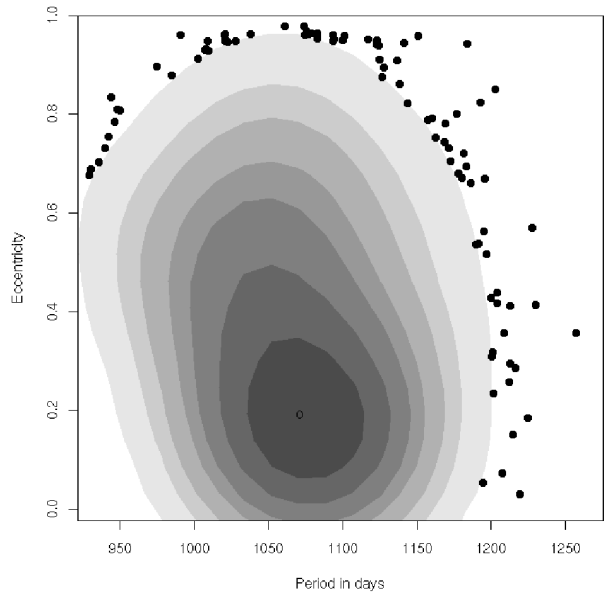}
 \caption{The joint posterior distribution of orbital period $T$ and eccentricity $e$ for the companion to HD 187085. The contours represent probability coverage=0.01,0.05,0.1,0.2,0.3,0.4,0.5 and 0.7. The circle in the centre represent the mode of the joint probability distribution. The dots outside the contour show the samples that fall outside 99\% coverage probability.}
 \label{fig:jointHD187085}
\end{figure}

Detection of a planet orbiting HD 187085 was announced in 2006 with 40
epoch of radial velocity measurements between 1998 November and 2005
October \citep{jones2006}. It has reported that although an orbital
solution of period 986 days and eccentricity 0.47 gave the best fit,
a solution with low eccentricity also produced similar fit to
the observed data. Application of \emph{ExoFit} revealed the posterior
density distribution orbital parameters as shown in
Fig.~\ref{fig:histHD187085}.
The MCMC values with iterations are shown in the user's manual.
It is evident from Fig.~\ref{fig:histHD187085} that the posterior
density of eccentricity is a heavily skewed . The joint
probability distributing of orbital period and eccentricity is shown
in Fig.~\ref{fig:jointHD187085} with two dimensional HDRs. From
these plots it is reasonable to conclude that the data clearly favours
a low eccentricity orbit.

\begin{figure}
 \centering
 \includegraphics[scale=0.6]{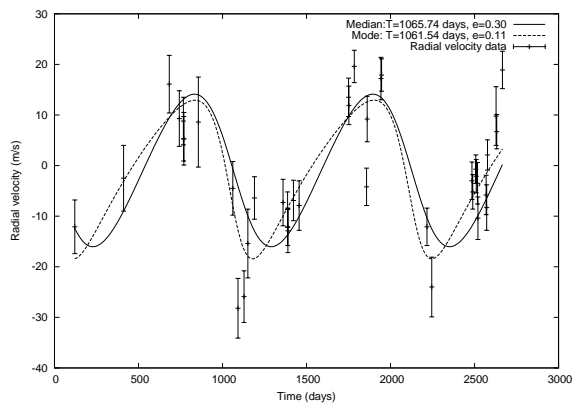}
 \caption{Possible orbital solutions for the companion to HD187085.}
 \label{fig:solsHD187085}
\end{figure}

Table~\ref{tab:solsHD187085} shows the summary statistics of the
posterior distribution of orbital parameters of HD 187085 along with
the published results. 
The eccentricity is $e=0.33, 0.30, 0.11$ from the mean, median and mode respectively,  compared with the published estimate $e=0.47$. \footnote{For a posterior distribution that is approximately Gaussian (e.g.. parameter T), estimates like mean, median and mode will yield nearly the same values. However, when the posterior density is not Gaussian and exhibit skewness (e.g.. parameter $e$) mean, median and mode will differ significantly. Since this is the case for parameter $e$, posterior median and and mode will be a better estimate than posterior mean. If the posterior distribution has more than one peak, posterior modes provide ideal summary of the distribution.}
Keplerian orbital solutions for HD 187085 from
Table~\ref{tab:solsHD187085} shown in
Fig.~\ref{fig:solsHD187085}.

How sensitive are the results to the assumed priors?
In Table~\ref{tab:10runs} we show results based on 
10 trial runs of for Gregory's priors (Table~\ref{tab:prior}) and for Top Hat priors,
i.e.  uniform between the same lower and upper values as in Gregory's priors. 
This indicates the data and assumed model are `good'.
The role of priors would be more dramatic in the case of poor data.
\begin{table*}
 \centering
 \caption{A comparison of estimators for orbital parameters of the companion to HD 187085.
The columns provide the 
posterior mean (and standard deviation), median (and 25\% and 75\%
quantiles) and the maximum a posteriori, i.e. posterior mode (and 68.3\% credible
regions). The last column is from the 
published results of Jones et al. (errors were  not provided in their paper). Here $m_p\sin\,i$ and $a$ are calculated by assuming the mass of the star $m_s=1.16\,M_{\odot}$.}
\begin{tabular}{lllll}
\hline
Parameter & \emph{ExoFit} & \emph{ExoFit} & \emph{ExoFit} & Jones et al  \\
\hline
 & Mean & Median & Mode & \\
\hline
$T(days)$ &$1065.84 \pm45.75 $ & $1065.74^{+30.48}_{-30.26}$ &
$1061.54^{+51.53}_{41.43}$ & 986 \\
\hline
$K(ms^{-1})$ &$16.55 \pm 4.49$  & $15.67^{+1.93}_{-1.49}$ &
$15.08^{+2.73}_{-2.28}$ & 17 \\
\hline
$V(ms^{-1})$ &  $ -0.94\pm1.56 $ & $-0.93^{+1.00}_{-1.02}$ &
$-0.81^{+1.38}_{-1.64}$ &  \\
\hline
$e$ & $ 0.33\pm 0.22$ &  $0.30^{+0.20}_{-0.15}$ & $0.11^{+0.29}_{-0.10}$ &
0.47 \\
\hline
$\varpi$ & $118.57 \pm23.19 $  & $112.39^{+18.58}_{-11.85}$ &
$95.32^{+29.22}_{-5.15}$ & 94 \\
\hline
$\chi$ & $0.11 \pm0.06 $  & $0.11^{+0.04}_{-0.04}$ & $0.10^{+0.07}_{-0.06}$
& \\
\hline
$s$ & $ 5.47\pm1.08 $  & $5.40^{+0.72}_{-0.68}$ & $5.21^{+1.15}_{-0.95}$ &
 \\
\hline
$T_p$ (JD -2450000)& $993\pm60.49 $  & $992^{+30}_{-32}$ &
$998.43^{+44.47}_{-64.29}$ & 912 \\
\hline
$m_p\sin i(M_{Jup})$ & $ 0.81\pm0.10 $  & $0.81^{+0.06}_{-0.07}$ &
$0.82^{+0.08}_{-0.12}$ & 0.75 \\
\hline
$a(AU)$ & $ 2.14\pm0.06 $  & $2.14^{+0.04}_{-0.04}$ & $2.14^{+0.06}_{-0.06}$
& 2.05\\
\hline
 \end{tabular}
 \label{tab:solsHD187085}
\end{table*}
\begin{table*}
 \centering
\caption{Mean and standard deviation of posterior modes of parameters  obtained from 10 trial
runs of $ExoFit$. }
\begin{tabular}{llllll}
\hline
 & HD 187085 & HD 187085 & &  HD159868 & HD159868  \\
 \hline
Parameter & Gregory's & Top Hat & $\qquad$  & Gregory's & Top Hat \\
\hline
T (days)& $1068.49\pm3.94$ & $1068.57\pm2.36$ & & $998.04\pm4.88$ & $1007.38\pm5.23$ \\
\hline
K $(ms^{-1})$ & $15.35\pm0.30$ & $16.34\pm0.66$ & & $43.16\pm4.93$ & $41.44\pm2.00$ \\
\hline
V $(ms^{-1})$& $-1.03\pm0.10$ & $-0.99\pm0.12$ & & $1.13\pm0.21$ & $1.05\pm0.28$ \\
\hline
e & $0.14\pm0.03$ & $0.16\pm0.51$ & & $0.70\pm0.04$ & $0.68\pm0.02$ \\
\hline
$\varpi$ & $96.30\pm2.86$ & $96.87\pm2.29$ & & $94.05\pm4.28$ & $98.50\pm3.28$ \\
\hline
$\chi$ & $0.11\pm0.004$ & $0.11\pm0.05$ & & $0.68\pm0.002$ & $0.68\pm0.001$ \\
\hline
s $(ms^{-1})$& $5.28\pm0.07$ & $5.03\pm0.06$ & & $9.52\pm0.10$ & $9.77\pm0.15$ \\
\hline
$T_p$ (JD -2450000) & $992.96\pm0.42$ & $993.74\pm0.05$ & & $706.15\pm2.77$ & $689.35\pm1.76$ \\
\hline
$m_p\sin\,i$ $(M_{Jup})$ & $0.80\pm0.006$ & $0.81\pm0.008$ & & $1.61\pm0.03$ & $1.60\pm0.01$ \\
\hline
a $(AU)$ & $2.14\pm0.005$ & $2.14\pm0.003$ & & $2.01\pm0.006$ & $2.02\pm0.006$\\
\hline
 \end{tabular}
 \label{tab:10runs}
\end{table*}

\subsection{The companion to HD 159868}
\begin{figure}
 \centering
 \includegraphics[scale=0.55]{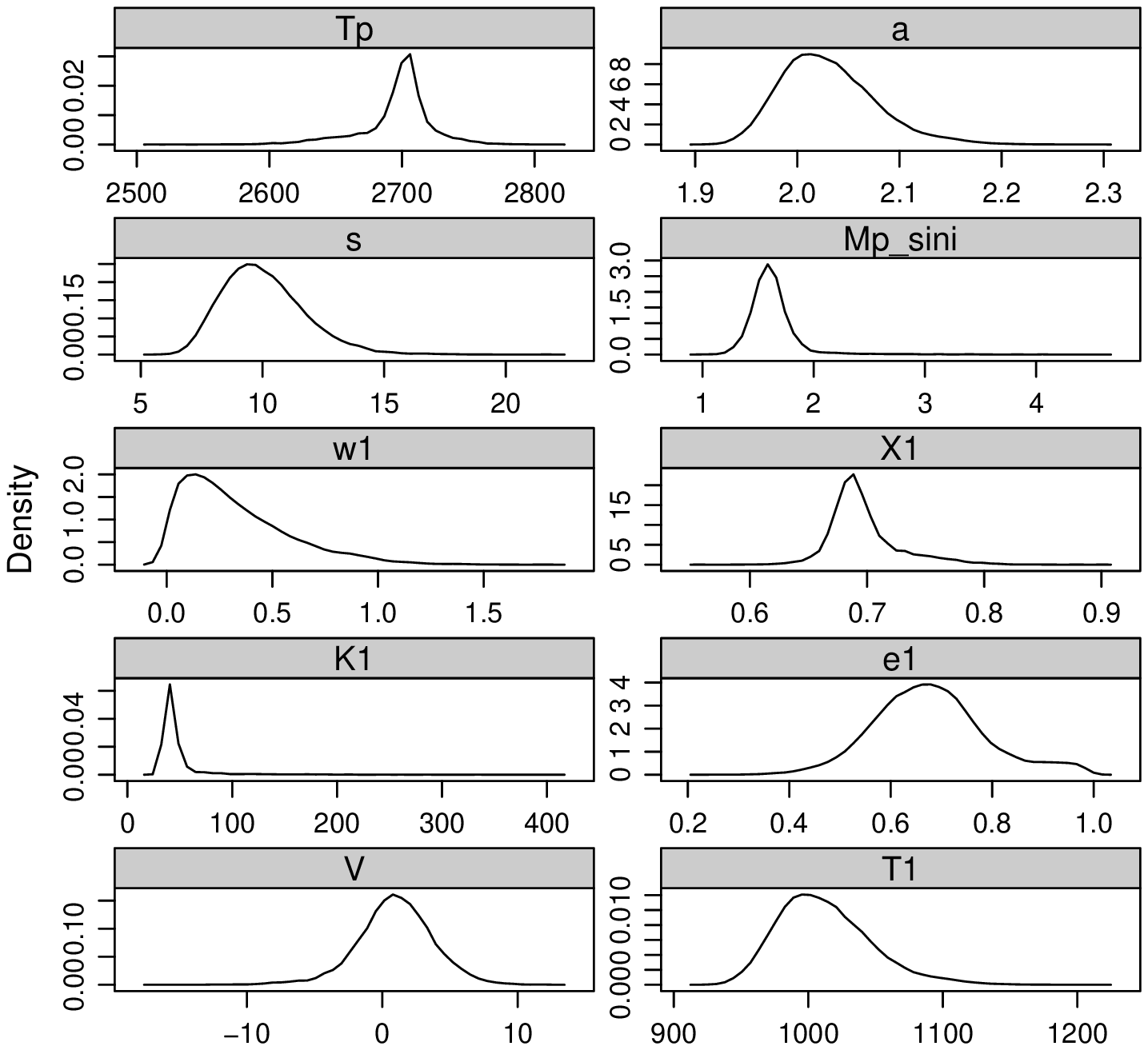}
 \caption{Posterior distribution of orbital parameters of the companion to HD 159868.}
 \label{fig:histHD159868}
\end{figure}

\begin{figure}
 \centering
\includegraphics[scale=0.55]{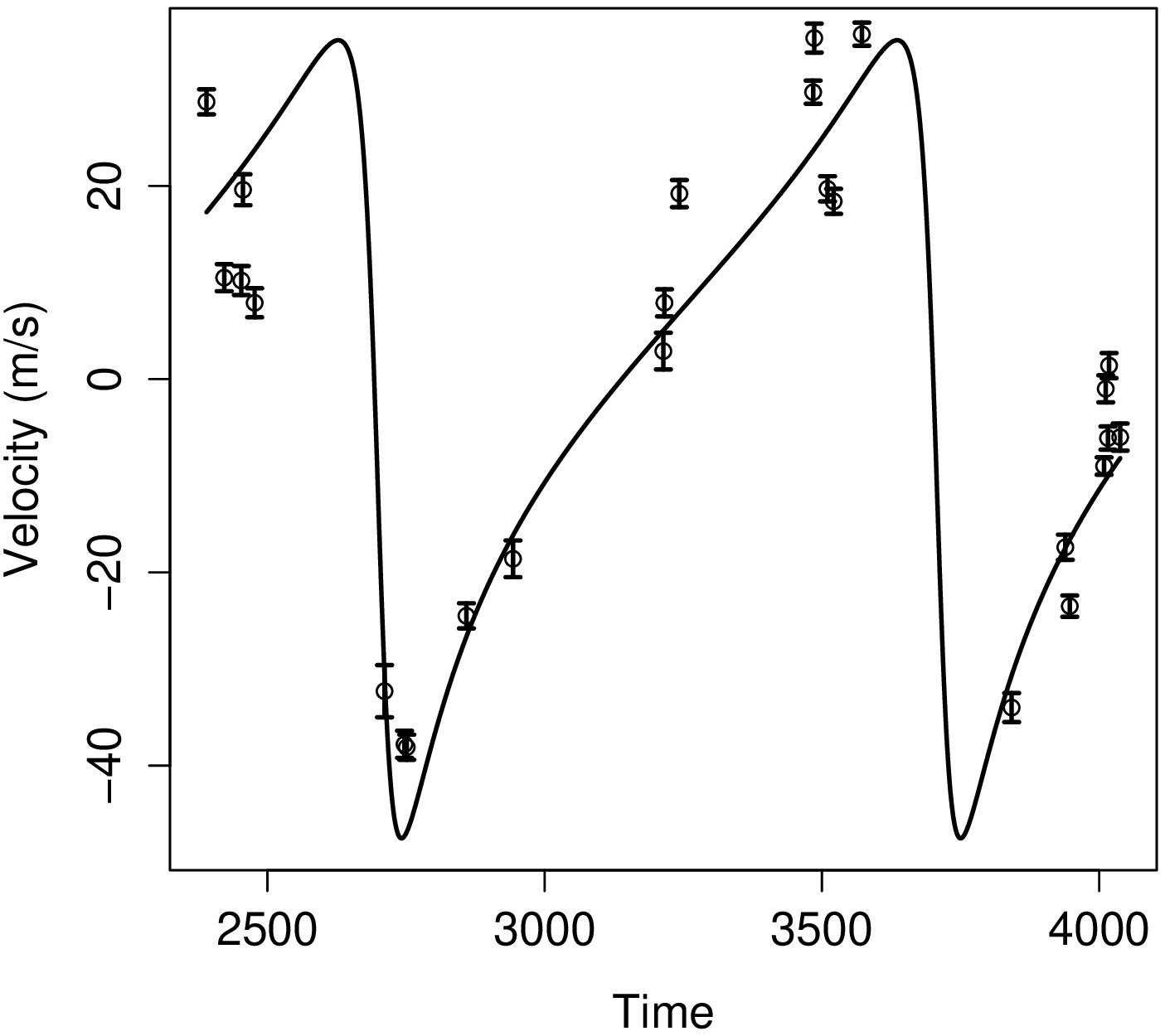}
 \caption{Best fit Keplerian orbit for the companion to HD 159868.}
 \label{fig:HD159868orbit}
\end{figure}

Orbital parameters of a planet orbiting HD 159868 from 28 radial
velocity measurements was reported on 2007 by \cite{OToole2007}. They
employed a two-dimensional periodogram to search for the optimum
orbital period and the eccentricity for the orbit. The posterior
distribution of parameters are shown in Fig.~\ref{fig:histHD159868}
and since they are nearly symmetric, we choose median of the
distribution as the point statistic. Table~\ref{tab:HD159856} shows
the median of samples from $ExoFit$ and the Keplerian orbit obtained
is plotted in Fig.~\ref{fig:HD159868orbit} along with the observed
data and uncertainties. Table~\ref{tab:10runs} shows the sensitivity
to changing the priors to Top Hat.  We note that the  `noise' factor $s$
is 10 times larger than the measurement errors, probably indicating
the presence of an unaccounted signal in the radial velocity data.

\begin{table*}
 \centering
 \caption{A comparison of the orbital parameters of the companion to HD 159868 from \emph{ExoFit} and the published results of \citet{OToole2007}. Here $m_p\sin\,i$ and $a$ are calculated by assuming the mass of the star $m_s=1.09\,M_{\odot}$. }
\begin{tabular}{|l|l|l|l|l|}
\hline
Parameter & \emph{ExoFit} & \emph{ExoFit} & \emph{ExoFit} & O'Toole et al  \\
\hline
 & Mean & Median & Mode & \\
\hline
T (days)&$ 1012.28\pm34.53$ & $1008.25^{+24.6}_{-20.71} $&$997.05^{+40.45}_{-24.41}$ & $986 \pm 9$ \\
\hline
K $(ms^{-1})$&$46.87\pm25.63$  &$ 41.30^{+5.17}_{-3.93}$ &$40.34^{+6.67}_{-6.6}$& $43.3 \pm 2$ \\
\hline
V $(ms^{-1})$&$0.96\pm 2.80$  &$ 1.00^{+1.77}_{-1.64}$ &$0.89^{+2.71}_{-2.37}$&  \\
\hline
e &$ 0.67\pm 0.11$ & $0.67^{+0.06}_{-0.07} $&$0.67^{+0.09}_{-0.11}$& $0.69 \pm 0.02$ \\
\hline
$\varpi$ &$108.33\pm14.32$  &$ 105.11^{+11.4}_{-7.44}$&$96.81^{+16.10}_{-16.44}$ & $97 \pm 3$ \\
\hline
$\chi$ &$0.69\pm 0.02$ &$ 0.69^{+0.01}_{-0.02} $&$0.68^{+0.02}_{-0.02}$& -- \\
\hline
s $(ms^{-1})$&$10.06\pm1.69$  & $9.86^{+1.18}_{-1.01}$&$9.49^{+1.71}_{-1.41}$ & -- \\
\hline
$T_p$ (JD -2450000) &$672.44\pm26.5$ & $684.89^{+9}_{-11} $&$710.80^{+14.35}_{-20.74}$& $700 \pm 9$ \\
\hline
$m_p\sin\,i$ $(M_{Jup})$&$ 1.62\pm0.23$  & $1.59^{+0.09}_{-0.09}$&$1.59^{+0.15}_{-0.14}$ & $1.7 \pm 0.3$ \\
\hline
a $(AU)$ &$2.03\pm0.04$ & $2.02^{+0.03}_{-0.028}$&$2.01^{+0.05}_{-0.04}$ & $2 \pm 0.3 $\\
\hline
 \end{tabular}

\label{tab:HD159856}
\end{table*}
\section{Discussion}
\label{sec:discussion}
 We present a new software package, \emph{ExoFit}, for estimating the
 orbital parameters from radial velocity data in a Bayesian framework
 by utilising Markov Chain Monte Carlo (MCMC) simulations with the
 Metropolis-Hastings algorithm. \emph{ExoFit} can search for one or two planets in a given radial velocity data and can easily be extended to search for more planets. We applied $ExoFit$ to simulated data sets to check the accuracy of the parameters extracted. As an illustration we
 re-analysed the orbital solution of companions to HD 187085 and HD 159868 from the
 published radial velocity data.  We confirm the degeneracy reported
 for orbital parameters of the companion to HD 187085 and show that a low eccentricity
 orbit is more probable for this planet. For HD 159868 we obtained
 slightly different orbital solution and a relatively high `noise'
 factor indicating the presence of an unaccounted signal in the radial
 velocity data.  We have also studied the sensitivity of the results
 to changes in the Bayesian priors. We plan to extend \emph{ExoFit} to search for more than two planets and to analyse transit photometry data.
\section{Acknowledgements}
We are grateful to Jean-Philippe Beaulieu, Farah Islam, David Kipping,
Yasushi Suto and Giovanna Tinetti for helpful discussions.  SB
acknowledges a Royal Astronomical Society studentship grant and OL
acknowledges a Royal Society Wolfson Research Merit Award.

\bibliographystyle{mn2e}
\bibliography{exofit18}
\label{lastpage}
\end{document}